\documentclass[]{emulateapj}
\shortauthors{TOYOUCHI \& CHIBA.}
\shorttitle{On the Chemical and Structural Evolution of the Galactic Disk}

\begin{document}

\title{On the Chemical and Structural Evolution of the Galactic Disk}

\author{Daisuke~Toyouchi\altaffilmark{1} and
	Masashi~Chiba\altaffilmark{1}}

\altaffiltext{1}{Astronomical Institute, Tohoku University,
Aoba-ku, Sendai 980-8578, Japan}

%%%%%% Abstract %%%%%%%%%%%%%%%%%%%%%%%%%%%%%%%%%%%%%%%%%%
\begin{abstract}
We study the detailed properties of the radial metallicity gradient in the stellar disk of our Galaxy to constrain its chemical and structural evolution. For this purpose we select and analyze $\sim$ 18,500 disk stars taken from two datasets, the Sloan Digital Sky Survey (SDSS) and the High-Accuracy Radial velocity Planetary Searcher (HARPS). On these surveys we examine the metallicity gradient, $\Delta$[Fe/H]/$\Delta R_{\rm g}$, along the guiding-center radii, $R_{\rm g}$, of stars and its dependence on the [$\alpha$/Fe] ratios, to infer the original metallicity distribution of the gas disk from which those stars formed and its time evolution. In both sample sources, the thick-disk candidate stars characterized by high [$\alpha$/Fe] ratios ([$\alpha$/Fe] $>$ 0.3 in SDSS, [$\alpha$/Fe] $>$ 0.2 in HARPS) are found to show a positive $\Delta$[Fe/H]/$\Delta R_{\rm g}$, whereas the thin-disk candidate stars characterized by lower [$\alpha$/Fe] ratios show a negative one. Furthermore, we find that the relatively young thin-disk population characterized by much lower [$\alpha$/Fe] ratios ([$\alpha$/Fe] $<$ 0.2 in SDSS, [$\alpha$/Fe] $<$ 0.1 in HARPS) shows notably a flattening $\Delta$[Fe/H]/$\Delta R_{\rm g}$ with decreasing [$\alpha$/Fe], in contrast to the old one with higher [$\alpha$/Fe] ratios ([$\alpha$/Fe] $\sim$ 0.2 in SDSS, [$\alpha$/Fe] $\sim$ 0.1 in HARPS). The possible implication for early disk evolution is discussed, in the context of galaxy formation accompanying the rapid infall of primordial gas on the inner disk region, which can generate a positive metallicity gradient, and the subsequent chemical evolution of the disk, which results in a flattening effect of a metallicity gradient at later epochs.

\end{abstract}
%%%%%%%%%%%%%%%%%%%%%%%%%%%%%%%%%%%%%%%%%%%%%%%%%%%%%%%%%%


%%% Sec.1 %%%%%%%%%%%%%%%%%%%%%%%%%%%%%%%%%%%%%%%%%%%%%%%%
\section{INTRODUCTION}
Our understanding of how disk galaxies like our own Galaxy have formed and developed their stellar disks is yet incomplete, although these galaxies constitute a major fraction in the present day of the Universe. The advent of recent high-resolution hydrodynamical simulations of galaxy formation in the cosmological context has highlighted basic baryonic processes therein, such as cold accretion flow of gas and energy feedback associated with supernovae explosion, which appear to play important roles in the formation of disk components. However, a grand picture for disk formation and evolution within hierarchical assembly of cold dark matter is yet unreached, where several unsolved issues reside in simulated disks including angular momentum problem (e.g., Navarro \& Steinmetz 2000) and other discrepancies compared with observed disks (e.g., Scannapieco et al. 2012).

In general, the structural and chemical evolution of a stellar disk is tightly related to past infall and redistribution of gas in the disk, which can affect subsequent star formation and chemical enrichment processes. Thus, disk formation history can be traced by the chemical abundances of stars distributed over a disk region. In particular, spatial gradients in the elemental abundances of disk stars are an important constraint on how a stellar disk has formed and evolved (e.g., Lacey \& Fall 1985; Lambert 1989; Goetz \& Koeppen 1992).

The radial metallicity gradient of the Galactic disk has been measured by using various tracers including HII regions, gaseous nebulae, star clusters and field stars. Cepheid variables are one of the good tracers for the radial metallicity gradient as the correct distance estimates are available (e.g., Andrievsky et al. 2002; Luck et al. 2006). The latest result using Cepheids shows a negative slope of $\Delta$[Fe/H]/$\Delta R$ $=$ $-$0.062 $\pm$ 0.002 dex kpc$^{-1}$ (Luck \& Lambert 2011). Since Cepheids are young populations with age $<$ 200 Myr, the radial metallicity gradient that they show is regarded as the present day gradient of the Galactic disk. The time evolution of the radial metallicity gradient can be probed by planetary nebulae and open clusters, which cover a larger range of ages than Cepheids. The studies using such tracers suggest that the radial metallicity gradient becomes slightly flatter with time (e.g., Friel et al. 2002; Chen et al. 2003, Maciel \& Costa 2009), although it is somewhat unclear due to the large uncertainty in the age determination. 

Similarly, studying the metallicity gradient using F, G, K dwarf disk stars, which include stars as old as the Galactic age, is useful to obtain important insights into the early chemical evolution process in the disk. Many previous observations of such stars have suggested the evidence for the thick-disk component, characterized by a more extended distribution from the disk plane, hotter kinematics, older age, lower [Fe/H] and higher [$\alpha$/Fe] than the main thin-disk component (e.g., Yoshii 1982; Gilmore \& Reid 1983; Schuster et al. 1993; Edvardsson et al. 1993; Wyse \& Gilmore 1995; Chiba \& Beers 2000; Robin et al. 2003; Soubiran et al. 2003; Bensby et al. 2003, 2005; Venn et al. 2004; Wyse et al. 2006; Reddy et al. 2006; Fuhrmann 2008; Juri\' c et al. 2008; Navarro et al. 2011)\footnote[1]{The question if these disks are physically separate components is discussed in the literature; Sch\"{o}nrich \& Binney (2009a), Loebman et al. (2011) and Bovy et al. (2012) showed that the observational findings provide no evidence for a separate thick disk in contrast to a continuous secular evolution.}. The radial metallicity gradient for this old thick disk has been measured and found to show a flat or somewhat positive metallicity gradient, whereas the young thin-disk star candidates show a negative metallicity gradient (e.g., Allende Prieto et al. 2006; Haywood 2008; Lee et al. 2011b; Carrell et al. 2012; Cheng et al. 2012). These results may imply some evolution of the radial metallicity gradient in the Galactic disk system, but more detailed properties of this time evolution, such as the systematic dependence of the radial metallicity gradient on time, are yet uncertain.

In this paper, we thus revisit this issue by investigating the time evolution of the radial metallicity gradient in the Galactic disk stars. For this purpose we use the [$\alpha$/Fe] ratios of stars as the indicators of their ages, which play the role of time clock due to different chemical contamination timing between Type Ia and II SNe (Matteucci \& Greggio 1986; Wyse \& Gilmore 1988). Here we analyze the disk candidate stars selected from two different datasets with refined abundance calibration for [Fe/H] and [$\alpha$/Fe]. We use the Sloan Digital Sky Survey (SDSS: York et al. 2000), in which the sample stars are widely distributed at the vertical distances from the Galactic plane of $|z|$ $\gtrsim$ 300 pc, i.e., higher than a typical scale height of the thin disk of $h_z \simeq 300$ pc, and the High-Accuracy Radial velocity Planetary Searcher (HARPS: Mayor et al. 2003), which provides precise spectroscopic data for nearby stars at $|z|$ $<$ 100 pc, i.e., lower than $h_z$. Thus, we obtain the general properties of the Galactic disk which are independent of observed regions and survey  volumes.

We note here that for old stellar populations the metallicity gradient along their observed current radii does not reflect the original radial distribution when they formed, because stars have increasing random motion on average with time due to, e.g., scattering by giant molecular clouds. Therefore, in this work we measure the metallicity gradient along the guiding-center radius, $R_{\rm g}$, of each star's orbit instead of the current radius to diminish this so-called ``blurring" impact (Sch\"{o}nrich \& Binney 2009a), although the ``churning" effect, with which stars may still migrate due to exchanges of angular momentum with bar/spiral structures, cannot be mitigated here.

This paper is organized as follows. In Section 2, we present our selection method to assemble the disk candidate stars from the SDSS and HARPS samples, and describe the method to calculate the position and orbit of each star. In Section 3, we show the properties of the radial metallicity gradient as a function of [$\alpha$/Fe] for our disk samples. Subsequently, we discuss the implications of our results on the chemical and structural evolution of the Galactic disk in Section 4. Finally our conclusions are drawn in Section 5.

%%% Sec.2 %%%%%%%%%%%%%%%%%%%%%%%%%%%%%%%%%%%%%%%%%%%%%%%%
\section{SELECTION OF DISK-STAR CANDIDATES}
%%% Sec.2.1 %%%%%%%%%%%%%%%%%%%%%%%%%%%%%%%%%%%%%%%%%%%%%%%
\subsection{SDSS/SEGUE G-dwarf sample}
SDSS (York et al. 2000) has obtained $u$, $g$, $r$, $i$ and $z$ magnitudes for many stars in the Galaxy, and the Sloan Extension for Galactic Understanding and Exploration (SEGUE; Yanny et al. 2009) is a low resolution ($R$ $\sim$ 2,000) spectroscopic follow-up survey of SDSS. From SDSS/SEGUE Data release 7 (DR7: Abazajian et al. 2009) we select SEGUE G-type dwarf stars as our sample with colors and magnitudes in the range $0.48 < (g - r)_0 < 0.55$ and  $r_0 < 20.2$, respectively, where all magnitudes are corrected for dust extinction using the reddening maps of Schlegel et al. (1998). The magnitude range for SEGUE is 14.0 $<$ $g$ $<$ 20.3, whereby almost all G dwarfs selected from this catalog are distributed at the vertical distance $|z|$ $>$ 300 pc away from the Galactic plane. For these stars, the stellar parameters such as effective temperature, $T_{\rm eff}$, surface gravity, log$\,g$, and metallicity, [Fe/H] were determined by the SEGUE Stellar Parameter Pipeline (SSPP: Lee et al. 2008a, 2008b; Allende Prieto et al. 2008; Smolinski et al. 2011). We set a surface gravity cut, log$\,g$ $>$ 4.2, to eliminate giant stars, when selecting our sample. Typical external errors in these parameters are 180 K in $T_{\rm eff}$, 0.25 dex in log$\,g$, and  0.23 dex in [Fe/H] (Smolinski et al. 2011). To investigate the evolution of the radial metallicity gradient in our sample, we use [$\alpha$/Fe] ratios as a substitute for stellar ages. According to Lee et al. (2011a), for stellar spectra with relatively high signal-to-noise ratios, S$/$N $>$ 20, a typical external error in [$\alpha$/Fe] is less than 0.1 dex. Here we adopt only stars with S$/$N $>$ 30 as our sample to ensure that chemical uncertainties give no impacts on our final results. However this criterion causes a lack of more metal-poor stars in our SDSS sample at larger distances from the Sun since in a fixed color range metal-poor dwarf stars are intrinsically fainter than metal-rich ones. In Section 4.1, we discuss the influence of this sampling bias on our results.

In order to determine distances from the Sun to individual stars, we adopt the relation shown in equation (A2), (A3) and (A7) in Ivezi\'c et al. (2008):
\begin{equation}
M_r(g-i, {\rm [Fe/H]}) = M^0_r(g-i) + \Delta M_r({\rm [Fe/H]}) \:,
\label{eq:eq1}
\end{equation}
where
\begin{eqnarray}
M^0_r(g-i) = -5.06 + 14.32(g-i) - 12.97(g-i)^2 \nonumber \\
+ 6.127(g-i)^3 - 1.267(g-i)^4 + 0.0967(g-i)^5 \:,
\label{eq:eq2}
\end{eqnarray}
\begin{equation}
\Delta M_r({\rm [Fe/H]}) = 4.50 - 1.11({\rm [Fe/H]}) -0.18({\rm [Fe/H]})^2\:.
\label{eq:eq3}
\end{equation}
These formulae were obtained by using 11 star clusters observed in SDSS in the metallicity range from +0.12 to $-$2.50, and uncertainty of distances derived from this relation was demonstrated to be about 10 to 15\% (Ivezi\'c et al. 2008). For more distant stars, this distance error becomes more significant in the estimates of full spatial velocities. Thus, to guarantee accurate space motions, we limit our sample stars with distance less than 3 kpc from the Sun.

%%% Sec.2.2 %%%%%%%%%%%%%%%%%%%%%%%%%%%%%%%%%%%%%%%%%%%%%%%
\subsection{HARPS sample}
HARPS (Mayor et al. 2003) has detected extra-solar planetary systems surrounding main-sequence stars from its high-precision radial-velocity measurements. Based on the extremely high-resolution spectra ($R$ $\sim$ 110,000) obtained in this program, Adibekyan et al. (2012) determined the elemental abundances of 1,111 F, G, and K dwarf stars with high accuracy, i.e., typical uncertainties in both of the [Fe/H] and [$\alpha$/Fe] values are less than 0.03 dex. The signal-to-noise ratios range from $\sim$ 20 to $\sim$ 2000 and are higher than 200 for 55~$\%$ of the sample. Accurate parallaxes and proper motions for these stars are available from the Hipparcos Catalogue (van Leeuwen 2007). For each star, the trigonometric parallax, $\pi$, obtained by Hipparcos, is reliable only for the stars near the solar neighborhood; we select the stars with parallax errors, $\sigma_{\pi}$, satisfying $\sigma_{\pi}/\pi$ $<$ 0.13 as our sample.

%%% Sec.2.3 %%%%%%%%%%%%%%%%%%%%%%%%%%%%%%%%%%%%%%%%%%%%%%%
\subsection{Calculation of space motions and guiding-center radii}
Once the solar position in the Galaxy ($R_{\odot}$), the circular velocity of the local standard of rest ($V_{\rm LSR}$), and the local solar motion ($U_{\odot}$, $V_{\odot}$, $W_{\odot}$) with respect to the LSR are given, we can derive each star's motion relative to the Galactic center from the information of the distance, radial velocity, and proper motion. We adopt  $R_{\odot}$ = 8.0 kpc, $V_{\rm LSR}$ = 220 km$\,$s$^{-1}$ and ($U_{\odot}$, $V_{\odot}$, $W_{\odot}$) = (11.1, 12.24, 7.25) km$\,$s$^{-1}$ (Sch\"{o}nrich et al. 2010). We note that adopting somewhat different values for these parameters, e.g., $R_{\odot}$ = 8.27 kpc, $V_{\rm LSR}$ = 238 km$\,$s$^{-1}$ and $U_{\odot}$ = 14.0 km$\,$s$^{-1}$ (Sch\"{o}nrich 2012), gives no particular modification to our main conclusion. Errors in calculated space motions for each star, mainly depending on errors in the distance, radial velocity and proper motion, are estimated from 500 Monte Carlo realizations of all relating parameters. Then, typical uncertainties in an estimated angular momentum perpendicular to the disk plane, $L_{\rm z}$, are found to be generally smaller than 10\% and 5\% in our samples selected from SDSS and HARPS, respectively.

Next, we determine each star's orbit by assuming the likely form of the Galactic gravitational potential. We adopt the oblate potential given in Law et al. (2005), which consists of a Miyamoto-Nagai disk (Miyamoto $\&$ Nagai 1975), a Hernquist model for the bulge (Hernquist 1990) and a logarithmic potential for the dark halo. The potentials of these subcomponents ($\phi_{\rm disk},\:\phi_{\rm bulge},\:\phi_{\rm halo}$) are given by
\begin{equation}
\phi_{\rm disk}=-\frac{GM_{\rm disk}}{\sqrt{R^2+\left(a+\sqrt{z^2+b^2}\right)^2}}\:,
\label{eq:eq4}
\end{equation}
\begin{equation}
\phi_{\rm bulge}=-\frac{GM_{\rm bulge}}{r+c}\:,
\label{eq:eq5}
\end{equation}
\begin{equation}
\phi_{\rm halo}=v_{\rm halo}^2\ln{[R^2+\left(z^2/q^2\right)+d^2]}\:,
\label{eq:eq6}
\end{equation}
where we set $M_{\rm disk}$ = $1.0\times10^{11}$M$_{\odot}$, $a$ = 6.5 kpc, $b$ = 0.26 kpc, $M_{\rm bulge}$ = $3.4\times10^{10}$M$_{\odot}$, $c$ = 0.7 kpc, $v_{\rm halo}$ = 121 km$\:$s$^{-1}$, $q$ = 0.9, and $d$ = 13 kpc. The circular velocity, $V_{\rm circ}$, derived from this potential yields $\simeq$ 220 km$\:$s$^{-1}$ at the solar position and is roughly constant at $R >$ 3 kpc. For our sample stars, we calculate their pericenters, $r_{\rm peri}$, apocenters, $r_{\rm apo}$, and maximum heights from the disk plane for their orbit, $Z_{\rm max}$. In addition, this Galactic potential is also used to derive a guiding-center radius, $R_{\rm g}$, for each star, where under an axisymmetric condition, a star satisfies
\begin{equation}
\left(\frac{\partial {\phi}}{\partial R}\right)_{R=R_{\rm g},|z|=0}=\;\;\frac{V^2_{\rm circ}}{R_{\rm g}}\;\;=\;\;\frac{L^2_{\rm z}}{R^3_{\rm g}} \:\:,
\label{eq:eq7}
\end{equation}
i.e., gravitational force is equal to centrifugal force due to rotational motion relative to the Galactic center and the orbital radius of a star with circular orbit coincides with $R_{\rm g}$. We regard this radius as a better proxy for the location within a gas disk at which each star originally formed than the observed current radius and expect that a star formed at $R_{\rm g}$ $>$ $R$ ($R_{\rm g}$ $<$ $R$) has moved inward (outward) to the current location, $R$, in the course of its orbital motion. In the next section, we will analyze the metallicity gradients along $R_{\rm g}$, $\Delta$[Fe/H]/$\Delta R_{\rm g}$, instead of the observed current radius $R$ for our disk-star sample. In order to infer the chemical evolution of a progenitor gas disk which has eventually turned to the current stellar disk, it is important to investigate metallicity distribution as a function of $R_{\rm g}$ rather than $R$, because the metallicity of each star reflects that of the original location where the star formed (Przybilla et al. 2008; Nieva \& Przybilla 2012).

However we note here that even $R_{\rm g}$ may not reflect the original birth radius of a star because each disk star may have changed its $R_{\rm g}$ through minor mergers of satellites (e.g., Quinn et al. 1993) or angular momentum exchanges with bar/spiral structures (e.g., Sellwood \& Binney 2002), implying that the original metallicity gradient may already be washed out and not be revealed by investigating current metallicity gradients along $R_{\rm g}$. We discuss this issue in Section 5.

%%% Sec.2.4 %%%%%%%%%%%%%%%%%%%%%%%%%%%%%%%%%%%%%%%%%%%%%%%
\subsection{Selection of disk stars}
Finally we exclude possible halo stars from our sample using the criteria of azimuthal velocity relative to the Galactic center $V_{\phi}$ $>$ 50 km$\:$s$^{-1}$, metallicity [Fe/H] $\ge$ $-$1.2, and orbital parameters $r_{\rm apo}$ $<$ 20 kpc and $Z_{\rm max}$ $<$ 3 kpc, in order to minimize halo contamination in our sample. We adopt these criteria following the sample selection of Lee et al. (2011b) and the results of Carollo et al. (2010), who found that the halo component is dominant at $Z_{\rm max}$ $>$ 5 kpc.

%%% Figure 1 %%%

\begin{figure}
\begin{center}
\figurenum{1}
\includegraphics[width=11cm,height=10cm]{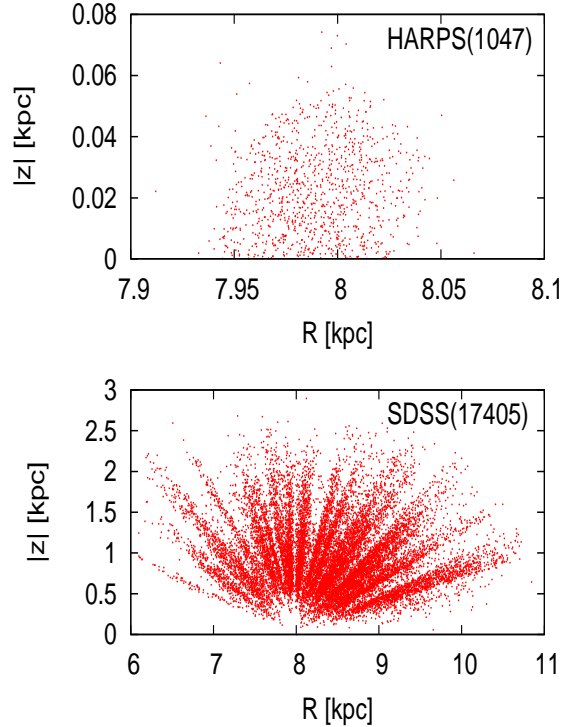}
\caption{Upper and lower panels show the spatial distributions of disk star candidates taken from the HARPS and SDSS samples, respectively. We assume that the Sun resides at ($R$,$\,|z|$) $=$ (8,$\,$0) kpc}
\label{fig:fig1} 
\end{center}
\end{figure}

Our final sample for disk star candidates includes 17,405 SDSS and 1,047 HARPS stars. Figure 1 shows the space distribution of our final sample. The HARPS stars are confined to $|z|$ $<$ 100 pc, whereas most of the SDSS stars reside at $|z|$ $>$ 300 pc and 6.5 kpc $<$ $R$ $<$ 10 kpc. Figure 2 shows the [Fe/H] vs. [$\alpha$/Fe] diagrams for the HARPS and SDSS samples. The stars are split into two distinct abundance trends separated at [$\alpha$/Fe] $\sim$ 0.2. These two distinct populations were also obtained in previous spectroscopic observations based on high-resolution spectra of nearby stars (Fuhrmann 1998; Bensby et al. 2003, 2005). It has been suggested that for the HARPS sample the division of the disk stars into two disk components based on the bimodal distribution of the [$\alpha$/Fe] ratios is broadly consistent with the kinematical decomposition between the thick and thin disk component (Adibekyan et al. 2012). This may imply that the transition from the thick to thin disk mostly occurs at [$\alpha$/Fe] $\sim$ 0.2. We thus phrase, in this paper, disk stars with higher (lower) [$\alpha$/Fe] than the ratio at this gap ([$\alpha$/Fe] $\sim 0.25$ in SDSS, [$\alpha$/Fe] $\sim 0.2$ in HARPS) as thick (thin) disk stars, which have been formed at the early (later) phase of disk formation\footnote[2]{It may be more appropriate to define such high (low) $\alpha$ disk stars as early (younger) disk populations rather than thick (thin) disk ones, since several recent studies, such as Bovy et al. (2012), reported no evidence of such distinct two-disk populations.}.

%%% Figure 2 %%%

\begin{figure}
\begin{center}
\figurenum{2}
\includegraphics[width=12cm,height=9cm]{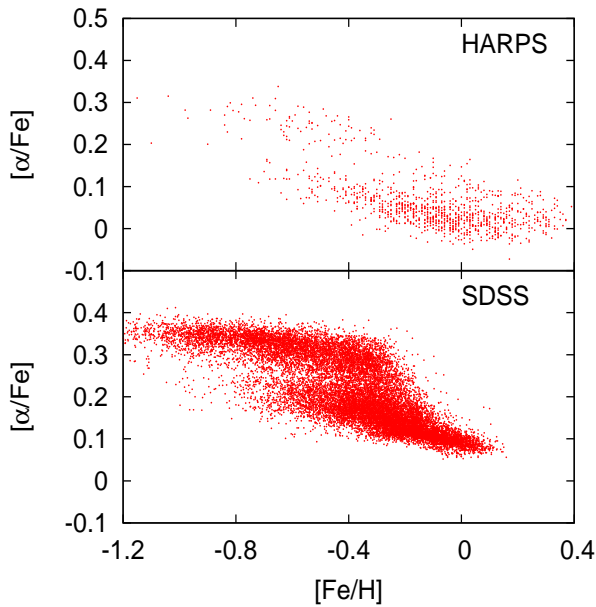}
\caption{Upper and lower panels show [Fe/H] vs. [$\alpha$/Fe] diagrams for the disk star candidates of the HARPS and SDSS stars, respectively.}
\label{fig:fig2} 
\end{center}
\end{figure}

%%% Sec.3 %%%%%%%%%%%%%%%%%%%%%%%%%%%%%%%%%%%%%%%%%%%%%%%%
\section{RESULTS}
%%% Sec.3.1 %%%%%%%%%%%%%%%%%%%%%%%%%%%%%%%%%%%%%%%%%%%%%%%   
\subsection{Metallicity gradients in the SDSS sample}
Figure 3 shows the [Fe/H] vs. $R_{\rm g}$ relation for the disk star candidates taken from our SDSS sample. In these panels, the stars are divided into two groups by the range of [$\alpha$/Fe], where dotted lines show a least-squares fitting to the unbinned samples. It is clear that $\Delta$[Fe/H]/$\Delta R_{\rm g}$ is positive for our thick disk stars with [$\alpha$/Fe] $>$ 0.3, compared with our thin disk stars with [$\alpha$/Fe] $<$ 0.2. These results are in agreement with the radial metallicity distribution as a function of a mean orbital radius, $R_{\rm mean}$, simply defined as $R_{\rm mean}$ $=$ ($R_{\rm max}\:+\:R_{\rm min}$)/2, where $R_{\rm max}$ ($R_{\rm min}$) is the maximum (minimum) distance projected on the disk plane from the Galactic center that a star reaches during its orbit (Nordstr\" om et al. 2004; Haywood 2008; Lee et al. 2011b; Co\c{s}kuno\u glu et al. 2012; Anders et al. 2013). This implies that the progenitor gas disk, which has subsequently formed the thick (thin) disk, may have experienced more (less) efficient chemical evolution in the outer disk region than the inner disk region. 

%%% Figure 3 %%%

\begin{figure}
\begin{center}
\figurenum{3}
\includegraphics[width=11cm,height=8cm]{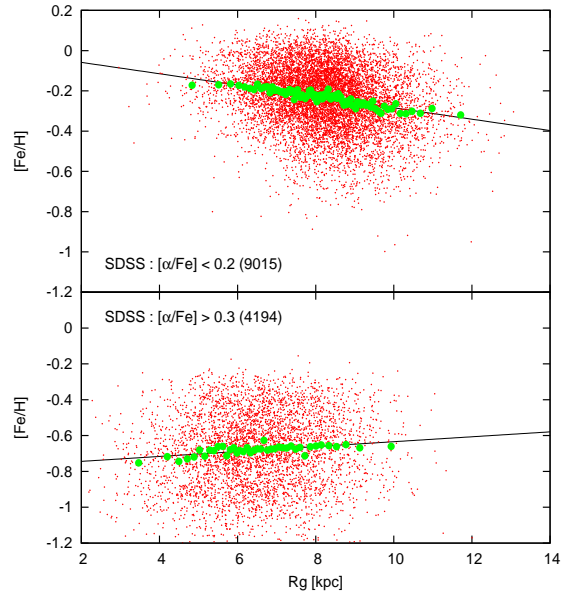}
\caption{Upper and lower panels show radial metallicity gradients along the guiding-center radii, $R_{\rm g}$ for the SDSS thin-disk candidates with [$\alpha$/Fe] $<$ 0.2 and thick-disk candidates with [$\alpha$/Fe] $>$ 0.3, respectively. Each green filled circle represents an average of 100 stars. The dotted line in each panel is obtained by a least-squares fitting to the unbinned stars.  The number of the sample stars is presented in each panel.}
\label{fig:fig3}
\end{center} 
\end{figure}

%%% Figure 4 %%%
\begin{figure*}
\begin{center}
\figurenum{4}
\includegraphics[width=17cm,height=13cm]{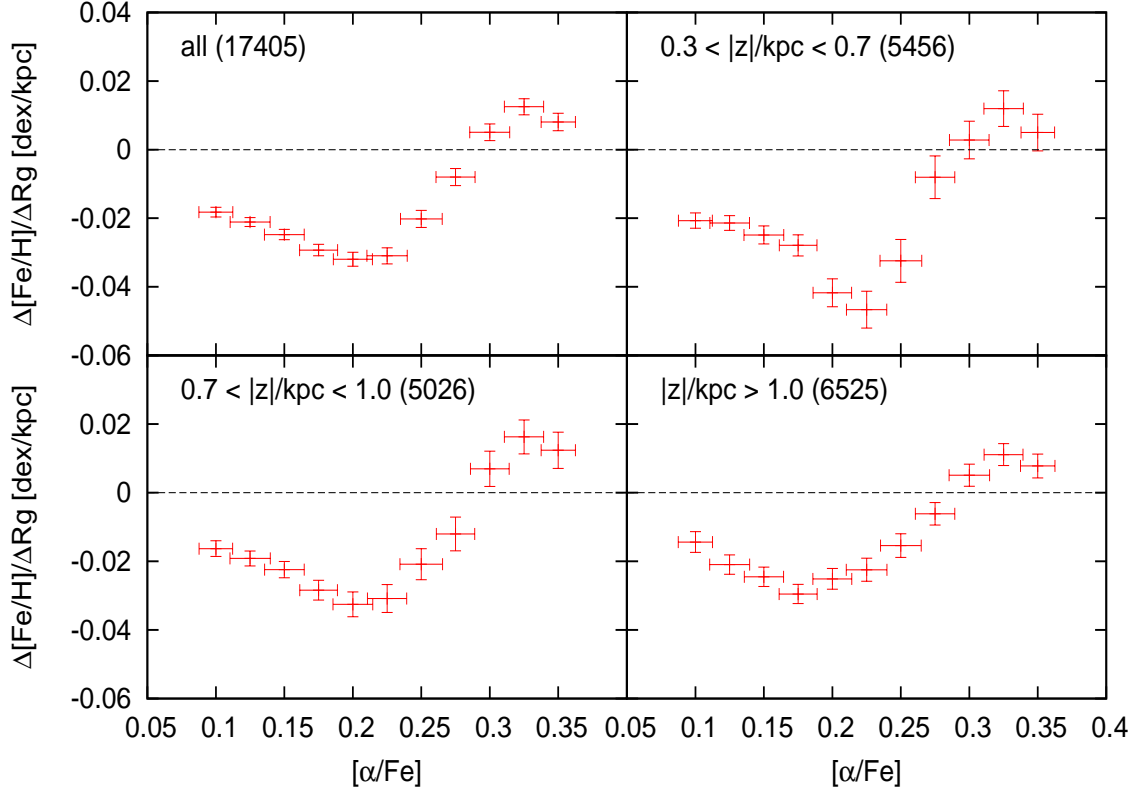}
\caption{The value of $\Delta$[Fe/H]/$\Delta R_{\rm g}$ as a function of [$\alpha$/Fe] for the SDSS sample. Each bin contains a sample of stars lying in $\pm$ 0.025 dex around any [$\alpha$/Fe] as a center of the bin, which moves by 0.025 dex towards larger [$\alpha$/Fe] in each step. The gradient of each bin is determined by a least-squares fitting to the unbinned stars clipped in each bin. Horizontal error bars represent the interval of [$\alpha$/Fe] for the clipped stars. Upper left panel shows the results for the all sample. Three other panels show the results for each group divided by the distance from the Galactic disk plane, $|z|$. The number of the sample stars is presented in each panel.}
\label{fig:fig4} 
\end{center}
\end{figure*}

We address the questions whether the metallicity gradient evolves continuously from the thick to thin disk formation phase and what its subsequent evolution over the course of the thin disk formation is. In order to obtain such further implications on the chemical evolution history of the Galactic disk, we divide our disk sample into finer bins in [$\alpha$/Fe] and investigate the dependence of $\Delta$[Fe/H]/$\Delta R_{\rm g}$ on [$\alpha$/Fe]. Upper left panel in Figure 4 shows $\Delta$[Fe/H]/$\Delta R_{\rm g}$ as a function of [$\alpha$/Fe] for all disk stars taken from our SDSS sample. In this figure, each bin contains stars lying in the interval of [$\alpha$/Fe] $\pm$ 0.025 dex, where the [$\alpha$/Fe] is its horizontal value, and this interval is moved by 0.025 dex towards larger [$\alpha$/Fe] in each step. As is clear, the metallicity gradient $\Delta$[Fe/H]/$\Delta R_{\rm g}$ is positive at [$\alpha$/Fe] $\gtrsim$ 0.3, then decreasing and becomes negative when [$\alpha$/Fe] decreases from 0.3 to 0.2, namely over the interval of [$\alpha$/Fe] separating the two distinct stellar populations as seen in Figure 2. This transition seems continuous, but it may be due to finite uncertainties in the [$\alpha$/Fe] estimation of the SDSS sample being as large as $\sim$ 0.1 dex. Another notable point is that for our thin disk stars the radial metallicity gradient becomes steepest at [$\alpha$/Fe] $\sim$ 0.2 and subsequently flatter with decreasing [$\alpha$/Fe] at [$\alpha$/Fe] $<$ 0.2. This implies that at the early phase after the transition from the thick to thin disk formation, the chemical enrichment in the inner disk region may have proceeded more efficiently than in the outer disk region, but the difference in the progression of chemical enrichment between the different radii is decreasing with time.

We also investigate the dependency on $|z|$ in the evolution of $\Delta$[Fe/H]/$\Delta R_{\rm g}$, for the SDSS sample stars are widely distributed from $\sim$ 300 pc to $\sim$ 2.0 kpc away from the Galactic plane. According to Lee et al. (2011b), the kinematical properties of the SDSS disk stars depend on the distance from the Galactic plane, i.e., the rotational velocities and orbital eccentricities of the disk stars decrease and increase with increasing $|z|$, respectively. This motivates us to check the effect of mixing stars over different ranges of $|z|$ on our results, because we use full kinematical information to calculate $R_{\rm g}$ for each star. Upper right and lower panels in Figure 4 show $\Delta$[Fe/H]/$\Delta R_{\rm g}$ as a function of [$\alpha$/Fe] for our SDSS disk stars divided into three groups in the range of $|z|$. We find that the general properties of the $\Delta$[Fe/H]/$\Delta R_{\rm g}$ vs. [$\alpha$/Fe] relation are similar for all panels, thereby indicating that the relation is general regardless of the difference of sampling in the vertical direction.

However, we note here that the positive $\Delta$[Fe/H]/$\Delta R_{\rm g}$ at [$\alpha$/Fe] $\gtrsim$ 0.3 and the steepest, negative slope at [$\alpha$/Fe] $\sim$ 0.2 may be artificial signatures caused by the contamination of stars with lower and higher [$\alpha$/Fe] stars, respectively, in each [$\alpha$/Fe] range, because of the finite uncertainty in [$\alpha$/Fe] measurements in SDSS/SEGUE. We discuss the influence of the contamination effect on our results in Section 4.2.

%%% Sec.3.2 %%%%%%%%%%%%%%%%%%%%%%%%%%%%%%%%%%%%%%%%%%%%%%%
\subsection{Metallicity gradients in the HARPS sample}
In order to assess whether the result obtained from the SDSS sample, which is mostly distributed at $|z|$ $>$ 300 pc, is truly universal, we perform the same analysis for the HARPS sample, in which stars are confined within 100 pc from the Sun. Figure 5 shows $\Delta$[Fe/H]/$\Delta R_{\rm g}$ as a function of [$\alpha$/Fe] for our HARPS disk stars. We find that the HARPS stars show the common properties for the evolution of $\Delta$[Fe/H]/$\Delta R_{\rm g}$ to the SDSS stars: the radial metallicity gradient of the Galactic disk was positive at the thick-disk formation phase, subsequently became negative during the transition from the thick to thin disk formation and was then made flatter with time during the thin disk formation phase. We conclude that these properties of $\Delta$[Fe/H]/$\Delta R_{\rm g}$ is a universal trend of the Galactic disk independent of distance from the disk plane.  The possible origin of these properties obtained here will be discussed in Section 4.3 and 4.4.

%%% Figure 5 %%%

\begin{figure}
\begin{center}
\figurenum{5}
\includegraphics[width=8cm,height=5.5cm]{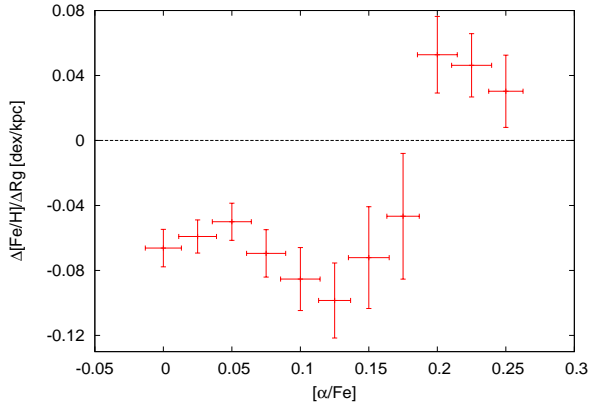}
\caption{The value of $\Delta$[Fe/H]/$\Delta R_{\rm g}$ as a function of [$\alpha$/Fe] for the HARPS sample selected in this work.}
\label{fig:fig5} 
\end{center}
\end{figure}

We note that there exist two minor differences between the SDSS and HARPS sample on the $\Delta$[Fe/H]/$\Delta R_{\rm g}$ vs. [$\alpha$/Fe] diagram. First, the transition of the sign of $\Delta$[Fe/H]/$\Delta R_{\rm g}$ for the HARPS sample appears rather discontinuous at [$\alpha$/Fe] $\sim$ 0.18 compared to the SDSS sample. Since the uncertainty in the [$\alpha$/Fe] estimation of the HARPS sample is much smaller than that of the SDSS sample, this discontinuous transition may be the case. Second, the [$\alpha$/Fe] ratio at the transition of the sign of $\Delta$[Fe/H]/$\Delta R_{\rm g}$ for the HARPS sample is about 0.1 dex smaller than that for the SDSS sample. This amount of difference in [$\alpha$/Fe] between both samples is also seen in the [$\alpha$/Fe] value of the gap in the [$\alpha$/Fe] vs. [Fe/H] diagram (see Figure 2). A possible reason for this difference in [$\alpha$/Fe] may originate from the different calibration method of this abundance ratio adopted in each survey. In fact, as mentioned by Venn et al. (2004, and references therein), it is likely that the different model atmospheres and atomic databases lead to a difference in the abundance ratios on the order of 0.1 to 0.2 dex.

%%% Sec.4 %%%%%%%%%%%%%%%%%%%%%%%%%%%%%%%%%%%%%%%%%%%%%%%%
\section{DISCUSSION} 
%%% Sec.4.1 %%%%%%%%%%%%%%%%%%%%%%%%%%%%%%%%%%%%%%%%%%%%%%%
\subsection{The influence of sampling bias on our results}
Before we introduce the possible origin of the properties shown in Section 3, we discuss how the effects of sampling bias in our disk samples influence our results. In particular, for the SDSS sample, there are finite selection biases for stars at both large and small distances from the Sun, which may largely affect the derived radial metallicity gradients. 

In the regions at large distances from the Sun, more metal-rich stars may be preferentially present in our sample selected with S/N $>$ 30. This is because such stars are intrinsically brighter than more metal-deficient stars due to the relation between star's luminosity and metallicity, and fainter stars are more likely to be discarded due to the adopted S/N cut. In Figure 6, to quantify this effect for our G-dwarf sample selected from SDSS/SEGUE, we show their $r$-band magnitudes vs. S/N relation. It is found that almost all stars brighter than $r_0$ = 16 mag meet our criterion of S/N $>$ 30. Then, using equation (1) combined with $(g - r)_0$ = 0.55 and [Fe/H] = $-$1.2 (the values obtained for the reddest and most metal-poor disk candidates, respectively, in our sample), this $r$-band magnitude corresponds to a distance, $d$, of 0.95 kpc from the Sun. Therefore, for the SDSS sample at $d <$ 0.95 kpc, the effect of the sampling bias due to the S/N cut on our result is expected to be negligible. 

%%% Figure 6 %%%

\begin{figure}
\begin{center}
\figurenum{6}
\includegraphics[width=8cm,height=5.5cm]{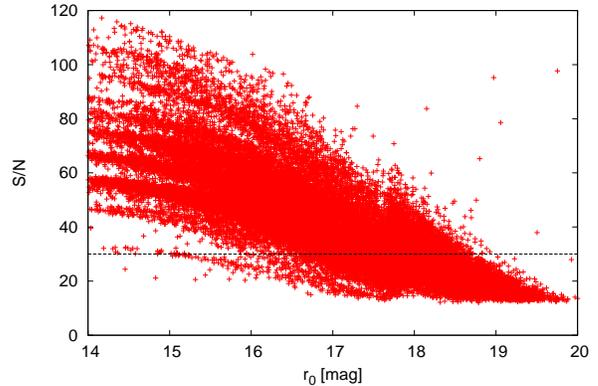}
\caption{$r$-band magnitude vs. S/N diagram for all G-dwarf stars observed in SDSS/SEGUE. The horizontal dashed line corresponds to our criterion of S/N $>$ 30.}
\label{fig:fig6} 
\end{center}
\end{figure}

%%% Figure 7 %%%

\begin{figure}
\begin{center}
\figurenum{7}
\includegraphics[width=8cm,height=5.5cm]{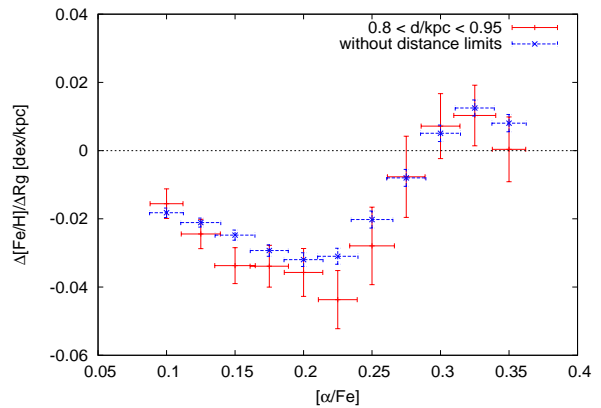}
\caption{The value of $\Delta$[Fe/H]/$\Delta R_{\rm g}$ as a function of [$\alpha$/Fe] for the SDSS sample stars in the distance range of 0.8 kpc $< d <$ 0.95 kpc (red), where the effect of the sampling biases on the result is expected to be negligible, and for those stars without distance limits (blue).}
\label{fig:fig7} 
\end{center}
\end{figure}

On the other hand, in the regions at small distances from the Sun, more metal-rich stars tend to be deficient in our sample since for such stars the distances corresponding to the brightest $r$-band magnitude, $r_0 \sim 14.0$, covered in SDSS/SEGUE are relatively larger than those for more metal-deficient stars. For the brightest star in our SDSS/SEGUE sample with $(g - r)_0$ = 0.48 and [Fe/H] = 0.2, we find the distance from the Sun of $\sim$ 0.8 kpc. Therefore, at $d >$ 0.8 kpc, we expect that our sample does not significantly suffer from the sampling bias due to the finite $r$-band magnitude range.

Following these discussion, we investigate the dependence of the radial metallicity gradient on [$\alpha$/Fe] for the SDSS/SEGUE disk stars in the distance range of 0.8 kpc $< d <$ 0.95 kpc, where both stars with high and low metallicities are well selected in our disk sample, and show the result in Figure 7. It is clear that the result for this distance range of stars is basically the same as that without distance limits.

We note here that there may exist a further sampling bias in SDSS/SEGUE due to the dependence of the selection function on $r$-band magnitudes (e.g., Bovy et al. 2012), but its effect would play a secondary role in our results, compared with the major sampling biases due to the S/N cut and observed $r$-band magnitude range. We thus conclude that the general properties of the radial metallicity gradient as a function of [$\alpha$/Fe] are not affected by selection effects in SDSS/SEGUE. As for HARPS, since the sample stars are located much closer to the Sun than for SDSS/SEGUE and no kinematical selection is made for our sampling, we expect that selection effects are minor.

%%% Sec.4.2 %%%%%%%%%%%%%%%%%%%%%%%%%%%%%%%%%%%%%%%%%%%%%%%
\subsection{On the contamination effects in this study}
As mentioned in Section 3.1, the positive metallicity gradient obtained for the disk stars with [$\alpha$/Fe] higher than the value of the gap in [$\alpha$/Fe] vs. [Fe/H] diagram may be caused by some contamination from stars with intrinsically lower [$\alpha$/Fe] due to uncertainty of [$\alpha$/Fe] measurements. For instance, the contamination of faster rotating (thus larger $R_{\rm g}$) population with low [$\alpha$/Fe], which preferentially occur at relatively high metallicities, can steepen the positive metallicity gradient. So we need carefully to assess if the positive gradient obtained in this study is an intrinsic property in the Galactic disk or not. For HARPS stars, this contamination effect is expected to be minor because the uncertainty in [$\alpha$/Fe] of 0.03 dex is smaller than the range covered by each bin in our Figure 5. However, for SDSS stars, the contamination effect would not be negligible due to the larger uncertainty in [$\alpha$/Fe] measurements than HARPS. We examine it by selecting more accurate data of S/N $>$ 50 for SDSS disk stars, in which the uncertainty in [$\alpha$/Fe] is confined to $\sigma$ $<$ 0.07 dex (Lee et al. 2011a). We have confirmed from this more accurate sample that high [$\alpha$/Fe] stars having [$\alpha$/Fe] $>$ 0.32 (i.e., at least 1 $\sigma$ higher than the value of [$\alpha$/Fe] = 0.25 dex corresponding to the observed gap in the [$\alpha$/Fe] vs. [Fe/H] distribution) indeed show a positive metallicity gradient. Thus, we conclude that our result of the positive metallicity gradient for high [$\alpha$/Fe] stars is not a contamination effect by [$\alpha$/Fe] values. 

At the intermediate [$\alpha$/Fe] ratios near the gap in the [$\alpha$/Fe] vs. [Fe/H] diagram, the contaminations of higher and lower [$\alpha$/Fe] populations, which show positive and negative radial metallicity gradients, respectively, are maximal. However, the observed metallicity gradient is expected to show not a steepening but flattening slope by such a contamination effect, since in this case the contamination included in a given population leads to weaken the amplitude of the slope in its intrinsic metallicity gradient. Thus, we believe that this contamination effect on the steepening slope with decreasing [$\alpha$/Fe], which occurs at the slightly lower [$\alpha$/Fe] than the value of the gap in the [$\alpha$/Fe] vs. [Fe/H] diagram, is also minor.

However, we need to remark the possibility of another contamination of disk populations with different age in a fixed [$\alpha$/Fe] slice, since the isochrone lines in the [$\alpha$/Fe] vs. [Fe/H] diagram are likely inclined. Particularly, such a contamination may affect the radial metallicity gradient at lower [$\alpha$/Fe] ratios than the gap in the [$\alpha$/Fe] vs. [Fe/H] diagram, where old metal-rich (young metal-poor) disk stars, which formed in the inner (outer) disk region, may be significantly mixed (see Figure 5 in Sch\"{o}nrich \& Binney 2009b). To what extent this mixture of different populations actually occurs in the concerned [$\alpha$/Fe] range depends on the assumptions and predictions of chemical evolution models. This issue is thus to be addressed and examined in chemo-dynamical models to reproduce the observed metallicity gradients and their dependence on [$\alpha$/Fe].

It is worth noting that as introduced below, the similar trends to Figure 4 and 5 are seen in the age dependence of radial metallicity gradients measured in Casagrande et al. (2011), thereby implying that the properties of the metallicity gradients obtained here are general. In the following subsections, we discuss the possible origin of these metallicity gradients in the context of the disk evolution.

%%% Sec.4.3 %%%%%%%%%%%%%%%%%%%%%%%%%%%%%%%%%%%%%%%%%%%%%%%
\subsection{The origin of the positive $\Delta$[Fe/H]/$\Delta R_{\rm g}$ in the thick disk}
Is the presence of the positive $\Delta$[Fe/H]/$\Delta R_{\rm g}$ in a protogalactic gas disk realistic in the context of disk galaxy formation and evolution$?$ This appears to be actually observed in gas-rich, young galaxies at high redshifts. Cresci et al. (2010) investigated the metallicity distribution of three rotationally-supported star formation galaxies at redshifts of $\sim$ 3 by examining the flux ratio between several rest-frame optical emission lines observed with VLT/SINFONI. They found that these galaxies show the positive radial gradient in their gaseous metallicty, and suggested that central gas in these galaxies may have been diluted by the infall of primordial gas, as expected by 'cold flow' galaxy formation models (e.g., Kere\v s et al. 2005; Dekel $\&$ Birnboim 2006; Kere\v s et al. 2009). More recently, Troncoso et al. (2013) reported further evidence of positive radial metallicity gradients in star-forming galaxies at redshifts of $\sim$ 3.4 using VLT/SINFONI. 

Such a gas disk with a positive metallicity gradient is actually achieved in some analytical chemical evolution models (e.g., Chiappini et al. 2001; Spitoni $\&$ Matteucci 2011; Mott et al. 2013): these models consider two infall episodes representing rapid and slow infall of primordial gas, which might form the thick disk and thin disk, respectively, whereas recently Minchev et al. (2013) suggested that the Galactic chemo-dynamical evolution model adopting only one infall episode can reproduce several observational results for the thick disk. It is worth noting that in these models, for the first a few Gyrs, the efficiency of chemical enrichment in the inner disk region is low due to the infall of a large amount of primordial gas, thereby resulting in the positive radial metallicity gradient of a gas disk. At a later epoch such a rapid gas inflow ceases and a metallicity gradient turns to be negative, as obtained for the thin disk in our Figure 4 and 5 (see Figure 11 of Chiappini et al. 2001). Since the most of thick disk stars are old and $\alpha$ enhanced, it is thought to have been rapidly built up during the first a few Gyrs at redshifts of 2 $\sim$ 3, which may correspond to the galaxies observed by Cresci et al. (2010) and Troncoso et al. (2013). On the other hand, the observed range of age and metallicity of the thin disk stars suggests that after the formation of the thick disk, the thin disk has gradually formed for an extended period of more than 6 Gyrs. Therefore the positive and negative $\Delta$[Fe/H]/$\Delta R_{\rm g}$ in the thick and thin disk, respectively, can be understood from the existing theoretical models.

We note that the positive $\Delta$[Fe/H]/$\Delta R_{\rm g}$ of the thick-disk stars can yield their positive $\Delta V_{\phi}$/$\Delta$[Fe/H], which implies that more metal-poor stars have slower azimuthal velocities relative to the Galactic center (e.g., Spagna et al. 2010; Lee et al. 2011b; Kordopatis et al. 2011; Adibekyan et al. 2013). This is understood based on the angular momentum conservation law, $L_{\rm z}$ $=$ $R$$V_{\phi}$ $=$ $constant$: for an axisymmetric system a star moving inward (outward) subsequently increases (decreases) $V_{\phi}$, so that for a stellar disk system formed from a gas disk with $\Delta$[Fe/H]/$\Delta R_{\rm g}$ $>$ 0, a star having moved from the inner region has lower [Fe/H] and smaller $V_{\phi}$, thereby giving the positive $\Delta V_{\phi}$/$\Delta$[Fe/H]. According to Curir et al. (2012), their N-body simulation of a disk system, which takes into account radial migration effects on stars, can reproduce the positive $\Delta V_{\phi}$/$\Delta$[Fe/H] of thick-disk stars if a positive radial metallicity gradient (as obtained in Spitoni \& Matteucci 2011) is adopted as an initial state of the disk. Thus, a rapid infall of primordial gas in the central disk region at an early epoch of the Galactic disk formation may be the origin of the positive radial metallicity gradient in the thick disk.

%%% Sec.4.4 %%%%%%%%%%%%%%%%%%%%%%%%%%%%%%%%%%%%%%%%%%%%%%%
\subsection{On the evolution of $\Delta$[Fe/H]/$\Delta R_{\rm g}$ in the thin disk population}
As presented in Section 3, the young thin-disk population (at [$\alpha$/Fe] $<$ 0.2 for SDSS and [$\alpha$/Fe] $<$ 0.1 for HARPS) shows a flatter $\Delta$[Fe/H]/$\Delta R_{\rm g}$ than the old thin-disk one. This accords with the results obtained by using the Galactic open clusters and planetary nebulae (e.g., Friel et al. 2002; Maciel $\&$ Costa 2009). This evolution of $\Delta$[Fe/H]/$\Delta R_{\rm g}$ may imply the inside-out process of the disk formation through the gas infall history with a shorter time scale at the inner disk region than the outer one. Indeed, some chemical evolution models taking into account such gas infall history yields the decrease of the radial metallicity gradient with time (Moll\' a et al. 1997; Portinari $\&$ Chiosi 1999; Hou et al. 2000). However the models considering the two infall episodes show the opposite result: their radial metallicity gradients become steeper with time. Therefore, such models showing a positive metallicity gradient at an early epoch, yet fail to reproduce our results for the young thin-disk population.

The inside-out process in the disk formation is also achieved by the radial flow in a gas disk due to, e.g., viscous accretion process (Silk \& Norman 1981). Namely for a rotationally supported gas disk, the transportation of angular momentum occurs between neighboring radii because of gravitational viscosity, and subsequently materials in inner regions move inward and those in outer regions move outward. As a result, more chemically enriched gas in inner regions spreads outwards and $\Delta$[Fe/H]/$\Delta R_{\rm g}$ gets flatter with time (e.g., Sommer-Larsen \& Yoshii 1989, 1990). The chemical evolution models in Thon \& Meusinger (1998) including such viscous accretion process show the flatter $\Delta$[Fe/H]/$\Delta R_{\rm g}$ at a later epoch, although they never show a positive radial metallicity gradient. The latter result may be due to the assumption of a just one-infall episode of primordial gas in their model, which does not properly take into account the formation process of the thick disk at an early epoch. Thus, further theoretical studies of chemical evolution in a galactic disk, in particular taking into account two-infall episodes in a viscous accretion disk, are needed to understand the properties of $\Delta$[Fe/H]/$\Delta R_{\rm g}$ obtained in this paper.

%%% Sec.4.5 %%%%%%%%%%%%%%%%%%%%%%%%%%%%%%%%%%%%%%%%%%%%%%%
\subsection{Comparing our results with the recent works}
We discuss the similar recent studies of a radial metallicity gradient in the Galactic disk stars and compare with our results. Boeche et al. (2013) studied the dependence of $\Delta$[Fe/H]/$\Delta R_{\rm g}$ on $Z_{\rm max}$ for the disk stars taken from the fourth data release of RAVE (Kordopatis et al. 2013) and Geneva-Copenhagen Survey (GCS: Nordstr\" om et al. 2004). Their results are mostly in agreement with ours: the radial metallicity gradients for their samples are negative at $Z_{\rm max}$ $<$ 0.5 kpc, which may be dominated by the thin disk stars, and become flatter and positive at $Z_{\rm max}$ $>$ 0.5 kpc, which may correspond to the dominance of the thick disk. However their disk stars show no further evolution in $\Delta$[Fe/H]/$\Delta R_{\rm g}$ of the thin disk component, as we discussed in the previous subsection, and a somewhat flat radial metallicity gradient even at high $Z_{\rm max}$, where the thick disk population dominates. These differences from our results may be due to the use of $Z_{\rm max}$ instead of [$\alpha$/Fe], where the latter is more correlated with stellar age.

Casagrande et al. (2011) determined age for each of the GCS disk stars using a reliable effective temperature estimated from the infrared flux method, and investigated the radial metallicity gradient as a function of age. Their results are remarkably consistent with ours (see their Figure 18), although they provided no further discussion on the properties of a radial metallicity gradient for the young thin-disk population showing a signature of a flatter gradient with younger age. Comparing with our results, [$\alpha$/Fe] $\sim$ 0.2 at which the radial metallicity gradient changes its sign corresponds to age of $\sim$ 8 Gyr, whereas [$\alpha$/Fe] $<$ 0.1 at which the gradient is flatter with decreasing [$\alpha$/Fe] is seen for the young thin-disk population with age of $<$ 4 Gyr. This relation between age and [$\alpha$/Fe]  should be confirmed together with the evolution of the radial metallicity gradient in chemical evolution models.

%%% Sec.5 %%%%%%%%%%%%%%%%%%%%%%%%%%%%%%%%%%%%%%%%%%%%%%%%
\section{SUMMARY AND CONCLUSIONS}
We have investigated the radial metallicity gradient of the Galactic stellar disk to constrain its chemical and structural evolution. For this purpose, we have analyzed the disk candidate stars taken from SDSS, in which the sample stars are distributed mostly at $|z|$ $>$ 300 pc, i.e., higher than a typical scale height of the thin disk, and HARPS, which provides precise spectroscopic data for nearby stars at $|z|$ $<$ 100 pc. Thus, our results based on these sample stars are not strongly biased in their vertical distribution. To infer the original metallicity distribution along the gas disk from which the stars formed and its time evolution, we have made use of the guiding-center radii, $R_{\rm g}$, of the stellar orbits to estimate the metallicity gradient, $\Delta$[Fe/H]/$\Delta R_{\rm g}$, and the dependence of $\Delta$[Fe/H]/$\Delta R_{\rm g}$, on the [$\alpha$/Fe] ratios of the stars. In both sample sources, the thick-disk candidate stars characterized by high [$\alpha$/Fe] ratios ([$\alpha$/Fe] $>$ 0.3 in SDSS, [$\alpha$/Fe] $>$ 0.2 in HARPS) have been found to show a positive $\Delta$[Fe/H]/$\Delta R_{\rm g}$, regardless of the distance from the disk plane, whereas the thin-disk candidate stars characterized by lower [$\alpha$/Fe] ratios show a negative slope. Furthermore, we have found that the relatively young thin-disk population characterized by much lower [$\alpha$/Fe] ratios ([$\alpha$/Fe] $<$ 0.2 in SDSS, [$\alpha$/Fe] $<$ 0.1 in HARPS) shows notably a flattening metallicity gradient with decreasing [$\alpha$/Fe] in contrast to the old one with higher [$\alpha$/Fe] ([$\alpha$/Fe] $\sim$ 0.2 in SDSS, [$\alpha$/Fe] $\sim$ 0.1 in HARPS). Comparing the evolution of the radial metallicity gradient with that obtained by existing chemical evolution models suggests the existence of rapid primordial gas infall on the central disk region at an early disk formation epoch, and the subsequent inside-out star formation and chemical enrichment processes.

In this work, we assume that the guiding center radius of each star is mostly conserved and regarded as the indicator of its birth radius. However, the dynamical events such as minor mergers of satellites or gravitational interactions between each star and bar/spiral structure can provide the systematic change for $R_{\rm g}$ of each star. This implies that the current $\Delta$[Fe/H]/$\Delta R_{\rm g}$ reflects not only the chemical evolution as discussed above but also the dynamical evolution relating to merging history or bar/spiral instability in the disk. Unfortunately, it is still unclear how these dynamical events influence current radial metallicity gradients for the Galactic disk. In fact, some numerical simulations suggest that such dynamical evolutions can wash out initial radial metallicity gradients of old stellar populations (Loebman et al. 2011), whereas other work shows opposite results (Curir et al. 2014). Our finding of a finite metallicity gradient in old stellar populations suggest that any dynamical mechanisms for mixing  effects are not strong enough to completely wash out an initial gradient. We thus believe that the evolution of the radial metallicity gradient of the Galactic disk as derived here can be used as an effective tool to constrain the chemo-dynamical evolution of the Galactic disk by combining with chemical evolution models taking into account the dynamical evolution such as in Sch\"{o}nrich \& Binney (2009a) and Minchev et al. (2013). This will be the next step in our study.

%%%%%%%%%%%%%%%%%%%%%%%%%%%%%%%%%%%%%%%%%%%%%%%%%%%%%%%%%
\acknowledgments
We are grateful for the referee for her/his constructive comments that have helped us to improve our paper substantially. This work has been supported in part by a Grant-in-Aid for Scientific Research (20340039, 18072001) of the Ministry of Education, Culture, Sports, Science and Technology in Japan and by JSPS Core-to-Core Program ``International Research Network for Dark Energy''.

%%%%%%%%%%%%%%%%%%%%%%%%%%%%%%%%%%%%%%%%%%%%%%%%%%%%%%%%%

%%%%%%%%%%%%%%%%%%%%%%%%%%%%%%%%%%%%%%%%%%%%%%%%%%%%%%%%%

\clearpage
%%%%%%%%%%%%%%%%%%%%%%%%%%%%%%%%%%%%%%%%%%%%%%%%%%%%%%%%%

%%%%%%%%%%%%%%%%%%%%%%%%%%%%%%%%%%%%%%%%%%%%%%%%%%%%%%%%%


\begin{thebibliography}{1}
\bibitem[Abazajian et al. (2009)]{aba}Abazajian, K. N., Adelman-McCarthy, J. K., Ag\" ueros, M. A., et al. 2009, \apjs, 182, 543
\bibitem[Adibekyan et al. (2013)]{adi13}Adibekyan, V. Z., Figueira, P., Santos, N. C., et al. 2013, \aap, 554, A44
\bibitem[Adibekyan et al. (2012)]{adi12}Adibekyan, V. Z., Sousa, S. G., Santos, N. C., et al. 2012, \aap, 545, A32
\bibitem[Allende Prieto et al. (2006)]{all06}Allende Prieto, C., Beers, T. C., Wilhelm, R., et al. 2006, \apj, 636, 804
\bibitem[Allende Prieto et al. (2008)]{all08}Allende Prieto, C., Sivarani, T., Beers, T. C., et al. 2008, \aj, 136, 2070
\bibitem[Anders et al. (2013)]{ande}Anders, F., Chiappini, C., Santiago, B.~X., et al.\ 2013, arXiv:1311.4549
\bibitem[Andrievsky et al. (2002)]{andr}Andrievsky, S. M., Kovtyukh, V. V., Luck, R. E., et al. 2002, \aap, 381, 32
\bibitem[Bensby et al. (2003)]{ben03}Bensby, T., Feltzing, S., \& Lundstr\" om, I. 2003, \aap, 410, 527
\bibitem[Bensby et al. (2005)]{ben05}Bensby, T., Feltzing, S., Lundstr\" om, I., \& Ilyin, I. 2005, \aap, 433, 185
\bibitem[Boeche et al. (2013)]{boe}Boeche, C., Siebert, A., Piffl, T., et al. 2013, \aap, 559, A59
\bibitem[Bovy et al. (2012)]{bov}Bovy, J., Rix, H.-W., Liu, C., et al. 2012, \apj, 753, 148
\bibitem[Carollo et al. (2010)]{caro}Carollo, D., Beers, T. C., Chiba, M., et al. 2010, \apj, 712, 692
\bibitem[Carrell et al. (2012)]{carr}Carrell, K., Chen, Y., \& Zhao, G. 2012, \aj, 144, 185
\bibitem[Casagrande et al. (2011)]{cas}Casagrande, L., Sch\" onrich, R., Asplund, M., et al. 2011, \aap, 530, A138
\bibitem[Chen et al. (2003)]{che03}Chen, L., Hou, J. L., \& Wang, J. J. 2003, \aj, 125, 1397
\bibitem[Cheng et al. (2012)]{che12}Cheng, J. Y., Rockosi, C. M., Morrison, H. L., et al. 2012, \apj, 746, 149 
\bibitem[Chiappini et al. (2001)]{chia}Chiappini, C., Matteucci, F., \& Romano, D. 2001, \apj, 554, 1044
\bibitem[Chiba \& Beers (2000)]{chib}Chiba, M., \& Beers, T. C. 2000, \aj, 119, 2843
\bibitem[Co\c{s}kuno\u glu et al. (2012)]{cos}Co\c{s}kuno\u glu, B., Ak, S., Bilir, S., et al. 2012, \mnras, 419, 2844
\bibitem[Cresci et al. (2010)]{cre}Cresci, G., Mannucci, F., Maiolino, R., et al. 2010, \nat, 467, 811
\bibitem[Curir et al. (2012)]{cur12}Curir, A., Lattanzi, M. G., Spagna, A., et al. 2012, \aap, 545, A133
\bibitem[Curir et al. (2014)]{cur14} Curir, A., Serra, A.~L., Spagna, A., et al.\ 2014, arXiv:1402.4976
\bibitem[Dekel \& Birnboim (2006)]{dek}Dekel, A., \& Birnboim, Y. 2006, \mnras, 368, 2
\bibitem[Edvardsson et al. (1993)]{edv}Edvardsson, B., Andersen, J., Gustafsson, B., et al. 1993, \aap, 275, 101
\bibitem[Friel et al. (2002)]{fri}Friel, E. D., Janes, K. A., Tavarez, M., et al. 2002, \aj, 124, 2693
\bibitem[Fuhrmann (1998)]{fuh98}Fuhrmann, K. 1998, \aap, 338, 161
\bibitem[Fuhrmann (2008)]{fuh08}Fuhrmann, K. 2008, \mnras, 384, 173
\bibitem[Gilmore \& Reid (1983)]{gil}Gilmore, G., \& Reid, N. 1983, \mnras, 202, 1025
\bibitem[Goetz \& Koeppen (1992)]{goe}Goetz, M., \& Koeppen, J.\ 1992, \aap, 262, 455
\bibitem[Haywood (2008)]{hay}Haywood, M. 2008, \mnras, 388, 1175
\bibitem[Hernquist (1990)]{her}Hernquist, L. 1990, \apj, 356, 359
\bibitem[Hou et al. (2000)]{hou}Hou, J. L., Prantzos, N., \& Boissier, S. 2000, \aap, 362, 921
\bibitem[Ivezi\' c et al. (2008)]{ive}Ivezi\' c, \v Z., Sesar, B., Juri\' c, M., et al. 2008, \apj, 684, 287
\bibitem[Juri\' c et al. (2008)]{jur}Juri\' c, M., Ivezi\' c, \v Z., Brooks, A., et al. 2008, \apj, 673, 864
\bibitem[Kere\v s et al. (2005)]{ker05}Kere\v s, D., Katz, N., Weinberg, D. H., \& Dav\' e, R. 2005, \mnras, 363, 2
\bibitem[Kere\v s et al. (2009)]{ker09}Kere\v s, D., Katz, N., Fardal, M., et al. 2009, \mnras, 395, 160
\bibitem[Kordopatis et al. (2013)]{kor13}Kordopatis, G., Gilmore, G., Steinmetz, M., et al. 2013, \aj, 146, 134
\bibitem[Kordopatis et al. (2011)]{kor11}Kordopatis, G., Recio-Blanco, A., de Laverny, P., et al. 2011, \aap, 535, A107
\bibitem[Lacey \& Fall (1985)]{lac}Lacey, C.~G., \& Fall, S.~M.\ 1985, \apj, 290, 154
\bibitem[Lambert (1985)]{lam}Lambert, D. L. 1989, in American Institute of Physics Conference Series, Vol. 183, Cosmic Abundances of Matter, ed. C. J. Waddington, 168$-$199
\bibitem[Law et al. (2005)]{law}Law, D. R., Johnston, K. V., \& Majewski, S. R. 2005, \apj, 619, 807
\bibitem[Lee et al. (2011a)]{lee11a}Lee, Y. S., Beers, T. C., Allende Prieto, C., et al. 2011a, \aj, 141, 90
\bibitem[Lee et al. (2011b)]{lee11b}Lee, Y. S., Beers, T. C., An, D., et al. 2011b, \apj, 738, 187
\bibitem[Lee et al. (2008a)]{lee08a}Lee, Y. S., Beers, T. C., Sivarani, T., et al. 2008a, \aj, 136, 2022
\bibitem[Lee et al. (2008b)]{lee08b}Lee, Y. S., Beers, T. C., Sivarani, T., et al. 2008b, \aj, 136, 2050
\bibitem[Loebman et al. (2011)]{loe}Loebman, S.~R., Ro{\v s}kar, R., Debattista, V.~P., et al.\ 2011, \apj, 737, 8
\bibitem[Luck et al. (2006)]{luc06}Luck, R. E., Kovtyukh, V. V., \& Andrievsky, S. M. 2006, \aj, 132, 902
\bibitem[Luck \& Lambert (2011)]{luc11}Luck, R. E., \& Lambert, D. L. 2011, \aj, 142, 136
\bibitem[Maciel \& Costa (2009)]{mac}Maciel, W. J., \& Costa, R. D. D. 2009, in IAU Symp., 254, The Galaxy Disk in Cosmological Context, ed. J. Andersen, J. Bland-Hawthorn, \& B. Nordstr\" om (Cambridge: Cambridge Univ. Press), 38P
\bibitem[Matteucci \& Greggio (1986)]{mat}Matteucci, F., \& Greggio, L. 1986, \aap, 154, 279
\bibitem[Mayor et al. (2003)]{may}Mayor, M., Pepe, F., Queloz, D., et al. 2003, The Messenger, 114, 20
\bibitem[Minchev et al. (2013)]{min}Minchev, I., Chiappini, C., \& Martig, M. 2013, \aap, 558, A9
\bibitem[Miyamoto et al. (1975)]{miy}Miyamoto, M., \& Nagai, R. 1975, \pasj, 27, 533
\bibitem[Moll\' a et al. (1997)]{mol}Moll\' a, M., Ferrini, F., D\' iaz, A. I. 1997, \apj, 475, 519
\bibitem[Mott et al. (2013)]{mot}Mott, A., Spitoni, E., \& Matteucci, F. 2013, \mnras, 435, 2918
\bibitem[Navarro et al. (2011)]{nav11}Navarro, J. F., Abadi, M. G., Venn, K. A., et al. 2011, \mnras, 412, 1203
\bibitem[Navarro \& Steinmetz (2000)]{nav00}Navarro, J. F., \& Steinmetz, M. 2000, \apj, 538, 477
\bibitem[Nieva \& Przybilla (2012)]{nie}Nieva, M.-F., \& Przybilla, N.\ 2012, \aap, 539, A143
\bibitem[Nordstr\" om et al. (2004)]{nor}Nordstr\" om, B., Mayor, M., Andersen, J., et al. 2004, \aap, 418, 989
\bibitem[Portinari \& Chiosi (1999)]{por}Portinari, L., \& Chiosi, C. 1999, \aap, 350, 827
\bibitem[Przybilla et al. (2008)]{prz}Przybilla, N., Nieva, M.-F., \& Butler, K.\ 2008, \apjl, 688, L103
\bibitem[Quinn et al. (1993)]{qui}Quinn, P. J., Hernquist, L., \& Fullagar, D. P. 1993, \apj, 403, 74
\bibitem[Reddy et al. (2006)]{red}Reddy, B. E., Lambert, D. L., \& Allende Prieto, C. 2006, \mnras, 367, 1329
\bibitem[Robin et al. (2003)]{rob}Robin, A. C., Reyl\' e, C., Derri\` ere, S., \& Picaud, S. 2003, \aap, 409, 523
\bibitem[Scannapieco et al. (2012)]{sca}Scannapieco, C., Wadepuhl, M., Parry, O. H., et al. 2012, \mnras, 423, 1726
\bibitem[Schlegel et al. (1998)]{sch}Schlegel, D. J., Finkbeiner, D. P., \& Davis, M. 1998, \apj, 500, 525
\bibitem[Sch\" onrich (2012)]{sch12}Sch\" onrich, R. 2012, \mnras, 427, 274
\bibitem[Sch\" onrich \& Binney (2009a)]{sch09a}Sch\" onrich, R., \& Binney, J. 2009a, \mnras, 396, 203
\bibitem[Sch\" onrich \& Binney (2009b)]{sch09b}Sch\" onrich, R., \& Binney, J.\ 2009b, \mnras, 399, 1145
\bibitem[Sch\" onrich et al. (2010)]{sch10}Sch\" onrich, R., Binney, J., \& Dehnen, W. 2010, \mnras, 403, 1829
\bibitem[Schuster et al. (1993)]{schu}Schuster, W. J., Parrao, L., \& Contreras Mart\' inez, M. E. 1993, \aaps, 97, 951
\bibitem[Sellwood \& Binney (2002)]{sel}Sellwood, J. A., \& Binney, J. J. 2002, \mnras, 336, 785 
\bibitem[Silk \& Norman (1981)]{sil}Silk, J., \& Norman, C. 1981, \apj, 247, 59
\bibitem[Smolinski et al. (2011)]{smo}Smolinski, J. P., Lee, Y. S., Beers, T. C., et al. 2011, \aj, 141, 89
\bibitem[Sommer-Larsen \& Yoshii (1989)]{som89}Sommer-Larsen, J., \& Yoshii, Y. 1989, \mnras, 238, 133
\bibitem[Sommer-Larsen \& Yoshii (1990)]{som90}Sommer-Larsen, J., \& Yoshii, Y. 1990, \mnras, 243, 468
\bibitem[Soubiran et al. (2003)]{sou}Soubiran, C., Bienaym\' e, O., \& Siebert, A. 2003, \aap, 398, 141
\bibitem[Spagna et al. (2010)]{spa}Spagna, A., Lattanzi, M. G., Re Fiorentin, P., \& Smart, R. L. 2010, \aap, 510, L4
\bibitem[Spitoni \& Matteucci (2011)]{spi}Spitoni, E., \& Matteucci, F. 2011, \aap, 531, A72
\bibitem[Thon \& Meusinger (1998)]{tho}Thon, R., \& Meusinger, H. 1998, \aap, 338, 413
\bibitem[Troncoso et al. (2014)]{tro}Troncoso, P., Maiolino, R., Sommariva, V., et al.\ 2014, \aap, 563, A58
\bibitem[van Leeuwen (2007)]{van}van Leeuwen, F. 2007, \aap, 474, 653
\bibitem[Venn et al. (2004)]{ven}Venn, K. A., Irwin, M., Shetrone, M. D., et al. 2004, \aj, 128, 1177
\bibitem[Wyse \& Gilmore (1988)]{wys88}Wyse, R. F. G., \& Gilmore, G. 1988, \aj, 95, 1404
\bibitem[Wyse \& Gilmore (1995)]{wys95}Wyse, R. F. G., \& Gilmore, G. 1995, \aj, 110, 2771
\bibitem[Wyse et al. (2006)]{wys06}Wyse, R. F. G., Gilmore, G., Norris, J. E., et al. 2006, \apj, 639, L13
\bibitem[Yanny et al. (2009)]{yan}Yanny, B., Rockosi, C., Newberg, H. J., et al. 2009, \aj, 137, 4377
\bibitem[York et al. (2000)]{yor}York, D. G., Adelman, J., Anderson, J. E., et al. 2000, \aj, 120, 1579
\bibitem[Yoshii (1982)]{yos}Yoshii, Y. 1982, \pasj, 34, 365
\end{thebibliography}
\end{document}